\documentclass{article}
\usepackage{smc2025}
\usepackage[caption=false, font=footnotesize]{subfig}
\usepackage{paralist}
\usepackage[figure,table]{hypcap}

\usepackage{booktabs,multirow}
\usepackage[acronym]{glossaries}

\newacronym{mir}{MIR}{Music Information Retrieval}
\newacronym{cnn}{CNN}{Convolutional Neural Network}
\newacronym{mae}{MAE}{Mean Absolute Error}
\newacronym{mse}{MSE}{Mean Squared Error}
\newacronym{bim}{BIM}{Basic Iterative Method}
\newacronym{fgsm}{FGSM}{Fast Gradient Sign Method}
\newacronym{pgd}{PGD}{Projected Gradient Descent}
\newacronym{nmf}{NMF}{Nonnegative Matrix Factorization}

\DeclareMathOperator{\sign}{sign}
\DeclareMathOperator{\clip}{clip}
\DeclareMathOperator{\Expect}{\mathbb{E}}
\newcommand{\Reals}{\mathbb{R}}
\newcommand{\cD}{\mathcal{D}}
\newcommand{\norm}[1]{\left\lVert#1\right\rVert}

\usepackage[english]{babel}
\usepackage{alphabeta}

\def\papertitle{Are Inherently Interpretable Models More Robust? \\
A Study in Music Emotion Recognition}

\author[1]{\mbox{\firstname{Katharina}\lastname{Hoedt}\email{katharina.hoedt@jku.at}}}
\author[1]{\mbox{\firstname{Arthur}\lastname{Flexer}\email{arthur.flexer@jku.at}}}
\author[1,2]{\mbox{\firstname{Gerhard}\lastname{Widmer}\email{gerhard.widmer@jku.at}}}

\affil[1]{\department{Institute of Computational Perception}\institution{Johannes Kepler University}\city{Linz}\country{Austria}\affiliationtype{University}}
\affil[2]{\department{LIT AI Lab}\institution{Linz Institute of Technology}\country{Austria}}

\completesetup

\title{\papertitle}
\begin{document}
	\capstartfalse
	\maketitle
	\capstarttrue

\begin{abstract}
	One of the desired key properties of deep learning models is the ability to generalise to unseen samples. When provided with new samples that are (perceptually) similar to one or more training samples, deep learning models are expected to produce correspondingly similar outputs. Models that succeed in predicting similar outputs for similar inputs are often called robust.
    Deep learning models, on the other hand, have been shown to be highly vulnerable to minor (adversarial) perturbations of the input, which manage to drastically change a model's output and simultaneously expose its reliance on spurious correlations. In this work, we investigate whether \textit{inherently interpretable deep models}, i.e., deep models that were designed to focus more on meaningful and interpretable features, are more robust to irrelevant perturbations in the data, compared to their black-box counterparts.
    We test our hypothesis by comparing the robustness of an interpretable and a black-box music emotion recognition model when challenged with adversarial examples. Furthermore, we include an adversarially trained model, which is optimised to be more robust, in the comparison. Our results indicate that inherently more interpretable emotion recognition models can indeed be more robust than their black-box counterparts, and achieve similar levels of robustness as adversarially trained models, at lower computational cost. 
\end{abstract}
	%
    
\section{Introduction}
\label{sec:Introduction}
In \gls*{mir}, as in any other application domain of machine learning, a central desideratum
is that learned models should generalise, that is, make correct predictions on new observations outside of the training data.
A special aspect of this is \textit{robustness}: the ability of a model
to make similar predictions when presented with similar inputs. In particular,
we expect a good model to be robust against irrelevant changes in its input.
However, it has previously been shown that most traditional, black-box deep learning
models are not robust in this sense.
Instead, they are vulnerable to imperceptible perturbations of the inputs~\cite{Szegedy2014IntruigingProperties}, which can drastically change the predictions and hence the performance.
While this issue was first discussed in the image domain~\cite{Szegedy2014IntruigingProperties}, it has similarly been found to be a relevant problem in other domains in which deep learning became dominant, such as speech~\cite{Carlini2018STT} and MIR~\cite{Kereliuk2015AudioAdversaries}. Generally, the relationship between (adversarial)
robustness and the ability to generalise has been shown to be a complex one~\cite{Stutz2019RobustnessAndGeneralization}.

In this work, we hypothesise that deep learning models that are \textit{inherently interpretable}, i.e., models that by design focus more on interpretable features, may also be inherently more robust against minor alterations of the input. Specifically, we look into a variant of a concept bottleneck model~\cite{Chowdhury2019MidLevel}. Here, predictions are not obtained directly from an input via complex computations, but instead are based on human-understandable concepts (here: so-called mid-level features), which are extracted with the help of a bottleneck layer. Based on these concepts, the final class predictions are obtained via a linear layer, resulting in more transparent model decisions as we can track which concept contributed to which extent towards a prediction.

We test the robustness of an inherently interpretable model subsequently by focusing on a specific task -- music emotion recognition from audio -- and using an adversarial attack as a probing tool. We compare the robustness of traditional, black-box models, of their adversarially trained versions, and of an inherently interpretable model from the recent MIR literature. As the model treats emotion recognition as a regression task, we also need to adapt an existing attack method to fit a regression setting.  
We use this adversarial attack to investigate to what extent it manages to distort the performance of the different types of models. Our results indicate that the inherently interpretable model is substantially more robust than a black-box model, and even exhibits similar robustness as models that have been deliberately adversarially trained.
This documents -- beyond the usefulness of interpretability \textit{per se} -- another advantage of inherently interpretable models in \gls*{mir}, calling for more research in this direction.

\section{Related Work}
\label{sec:Related Work}
\textbf{Linking Interpretability and Robustness}. In what follows, we hypothesise about a connection between interpretability and adversarial robustness. Previously, it was shown that robust models lead to explanations (here: saliency maps) that are easier to comprehend than non-robust models (e.g.,~\cite{Etmann2019RobustnessSaliencyInterpretability}). An idea similar to ours, yet with a focus on post-hoc interpretability, is an adversarial defence based on the promotion of model interpretations that are robust to adversarial perturbations~\cite{Boopathy2020InterpretabilityHelpsRobustness}. Also, different studies observed that models trained to present interpretable gradients (e.g., via some type of regularisation~\cite{Noack2021IntAdvRelationStudy} or saliency-guided training~\cite{Guesmi2024SaliencyGuidedRobustness}) result in more robust models. Here, the main difference from our hypothesis is that we do not require interpretable gradients, but look at models that were designed to focus on human-interpretable features during training. 
Somewhat related, input modifications have been used to gain insights about the learned embedding space of deep \gls*{mir} models~\cite{Kim2019NeighborTestsMIR}. The work most similar to ours investigates the same hypothesis as we argue for, and look into the robustness of an inherently interpretable deep image classifier~\cite{Rasheed2024ConBottleneckAdvRobust}. The setup is similar to ours, as the robustness of a standard (black-box) \gls*{cnn}, an inherently interpretable model, and adversarially trained variants are compared, but in the image domain. Our work can therefore be seen as an extension of these tests to a different domain.

\textbf{Inherently Interpretable Deep Learning Models}. 
The topic of inherently interpretable deep models is still mostly in early stages of development in various domains. In the image domain, prototype-based models (e.g.,~\cite{Chen2019ThisLooksLike}) were introduced as systems that learn to represent prototypes (i.e., (parts of) representative inputs of a class), and base their predictions on these representations. Related, concept bottleneck models~\cite{Koh2020ConceptBottleneck} try to extract (human-interpretable) concepts from input data, on which the final predictions are based. Concepts can here be predefined and learned, e.g., by using corresponding annotations if available~\cite{Koh2020ConceptBottleneck}, or extracted automatically, e.g., by learning concepts in a self-supervised way during training~\cite{Wang2023BotCL}. The original idea of concept bottleneck models~\cite{Koh2020ConceptBottleneck} is very similar to the model we will look at in our experiments~\cite{Chowdhury2019MidLevel}. 

The work we build upon in this paper was one of the first approaches in \gls*{mir} where a model was designed so that its predictions relied on human-understandable concepts~\cite{Chowdhury2019MidLevel}. Next to this, systems have been proposed that base their predictions on automatically learned prototypes, or meaningful transformations thereof~\cite{Zinemanas2021APNet,Loiseau2022PlayablePrototypes}. A different approach previously proposed is the idea to train a classifier and an interpreter simultaneously to develop an interpretable system~\cite{Parekh2024NMFInherentInterpretabilityAudio}. Another avenue investigated in \gls*{mir} is the use of (self-)attention for interpretability purposes, e.g., \cite{Won2019InterpretableMusicTagging}, but the assumptions underlying this type of approach have been questioned~\cite{Jain2019AttentionNotExplanation}. Beyond the scope of classification, a deep music generation system that is controllable via interpretable latent factors (chord and texture) was proposed in the past~\cite{Wang2020InterpretableMusicGeneration}.

\textbf{Adversarial Attacks in Regression Tasks}. 
Most of the literature concerning adversarial attacks in any domain is focused on systems trying to solve a classification task, e.g., image classification, emotion classification, or tasks like speech-to-text~\cite{Goodfellow2015FGSM,Carlini2018STT,Madry2018AdversarialTraining,Sturm2014DetermineHorse,Kereliuk2015AudioAdversaries}. Significantly less work covers regression tasks. One idea is to take an existing classification attack (e.g., \gls*{fgsm}~\cite{Goodfellow2015FGSM}, \gls*{pgd}~\cite{Madry2018AdversarialTraining} or \gls*{bim}~\cite{Kurakin2017BIM}) and use it to maximise the error between a system's output and its target~\cite{Nguyen2018RegressionAttackFGSMPGD,Mode2020AttackMultivariateTime}. In our subsequent experiments, we follow a similar approach, and apply \gls*{bim}~\cite{Kurakin2017BIM} to maximise the prediction error of an emotion recognition system. Other approaches include using a sensitivity analysis to formulate an adversarial regression attack~\cite{Balda2019AdvsClassificationRegression}, or utilising the Jacobian of the function learned by a system to compute adversarial perturbations~\cite{Gupta2021RegressionAttack}. 

\textbf{Adversarial Defences}. To make systems more robust against adversarial input perturbations, various methods have been proposed. They often aim at either trying to detect these adversarial perturbations, or reduce them~\cite{Huang2020DefenceSurvey}. One of the most notable methods is \emph{adversarial training}~\cite{Madry2018AdversarialTraining}, during which adversarially perturbed input samples are included in the training routine of a model, to improve its robustness. Due to the high computational complexity of adversarial training, also various adaptations were proposed, e.g., with a particular focus on efficiency~\cite{Zhang2019YOPO,Shafahi2019FreeAdversarialTraining}. In the field of \gls*{mir}, there has also been some early work on adversarial training~\cite{Kereliuk2015AdvTrainingMIR}. In this work, we will use an adversarially trained system as a baseline regarding the robustness against adversarial attacks. 

\section{Methods}
\label{sec:Methods}
In this section, we elaborate on the methods we use to investigate our hypothesis. The focus is a model proposed for the task of emotion recognition~\cite{Chowdhury2019MidLevel}, which uses a bottleneck to learn human-interpretable (``mid-level'') features, before deriving the final emotion prediction. We consider this to be an ``inherently interpretable MIR model" because its very design --- specifically, the bottleneck structure that forces the model to encode its predictions in terms of human-interpretable concepts --- is targeted towards interpretability. In the following paragraphs, we briefly summarise the data used to train and test this model, the model itself, and some variations thereof. We also discuss the adversarial attack with which we test the robustness of the investigated models, as well as adversarial training.

\subsection{The Data}
\label{subsubsec:Emotion Classifier:Data}
The emotion recognition system by Chowdhury et al.~\cite{Chowdhury2019MidLevel} was trained on two different datasets. The first one, \emph{Soundtracks (Stimulus Set 1)}~\cite{Eerola2011EmotionData}, contains 360 excerpts originating from 110 movie soundtracks. The annotations consist of expert ratings for five discrete emotion categories (anger, fear, sadness, happiness, and tenderness) and three additional categories following a dimensional emotion model (valence, energy, tension). All eight emotion annotations are subsequently used as targets to solve multiple (i.e., eight) regression tasks. The ratings are real numbers and range from 1.0 to 7.83. The song excerpts have an average length of $\approx$ 17 seconds. The Soundtracks dataset contains the core emotion information that we want to learn from, which is why it is used to train all model variants that are discussed subsequently. 

The second dataset is the \emph{Mid-Level Features Dataset}~\cite{Aljanaki2018MidLevelData}, which includes human-interpretable feature annotations by means of seven mid-level descriptors (melodiousness, articulation, rhythmic and tonal stability, rhythmic complexity, dissonance, and modality / majorness). It contains ratings by musicians for each mid-level descriptor (range: 1 to 10) for a total of 5,000 song snippets ($\approx$ 15 seconds). This dataset is subsequently used to train a bottleneck on human-interpretable feature annotations within the interpretable deep model.

Note that all 360 song excerpts of the Soundtracks dataset are also included in the Mid-Level Features dataset. Therefore, the mid-level ground-truth annotations are available for all song excerpts that we try to predict emotions for. 

The inputs of the recognition systems are logarithmic-scaled spectrograms. To compute them, we follow the preprocessing steps proposed by Chowdhury et al.~\cite{Chowdhury2019MidLevel}. First, audio is resampled to 22.05 kHz, and converted to a mono signal. Then, we choose 10 arbitrary seconds for each audio excerpt. From these, we compute spectrograms (frame size = 2048 and hop size = 705, leading to 31.25 frames per second, number of filters per octave = 24) which are amplitude-normalised before logarithmic scaling. 

\subsection{The Recognition Systems}
\label{subsubsec:Emotion Classifer:Model}
Chowdhury et al. propose multiple system variants, which are all based on a VGG-style \gls*{cnn} architecture~\cite{Chowdhury2019MidLevel}. The authors compare systems where emotions are predicted directly from the input spectrogram, and more interpretable counterparts, in which mid-level features are predicted first. For this type of model, the final emotion prediction is derived from the predicted mid-level features with a linear layer, maintaining a more transparent prediction decision. 

\begin{figure}[t]
 \centering
 \includegraphics[width=0.5\columnwidth,trim={5cm 8cm 0 8cm},clip]{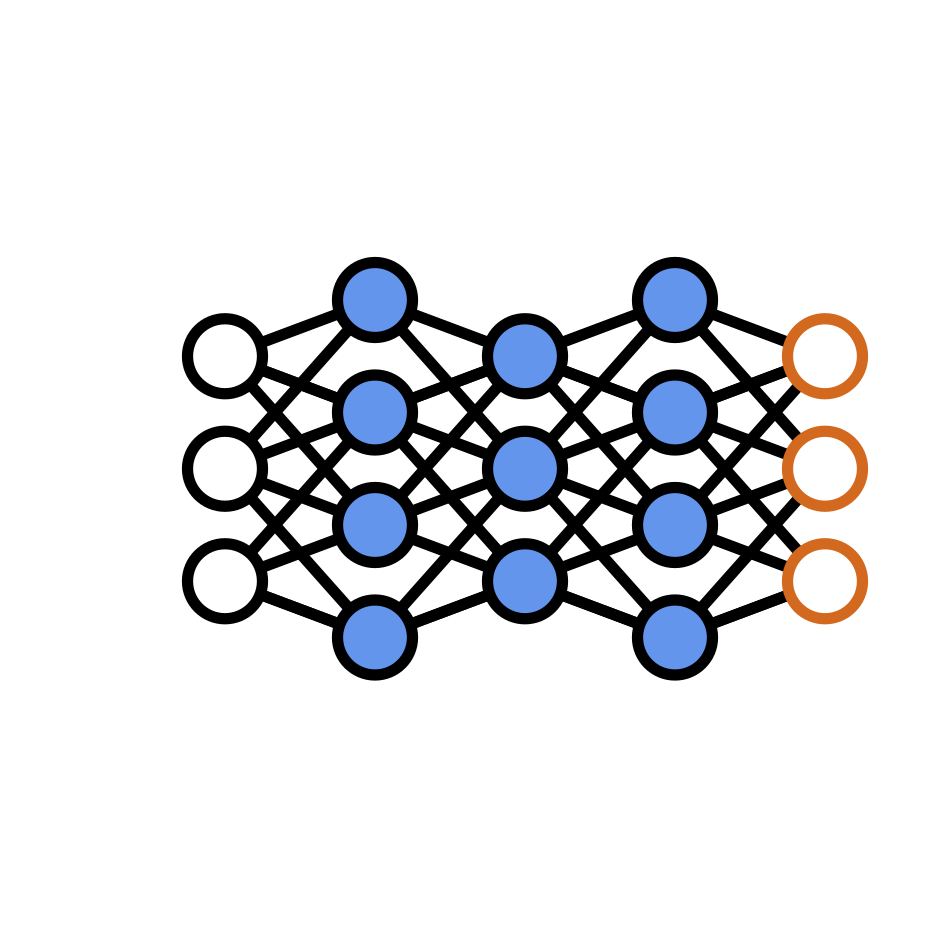}
 \caption{Schematic \emph{A2E} (black-box) model. During training, only output is optimised to predict emotions.}
 \label{fig:a2e}
\end{figure}

\textbf{A2E.} The first system we look at is the ``A2E'' variant~\cite{Chowdhury2019MidLevel}. This represents a traditional, black-box model that predicts emotions for the input directly based on (logarithmically scaled) input spectrograms. During training, each input is fed through multiple convolutional and pooling layers, and the output is optimised to predict eight different emotions. This model variant is trained on the Soundtracks dataset, and a \gls*{mse} loss is used to optimise for the emotion annotations. \figref{fig:a2e} depicts a schematic representation of this type of model. 

\begin{figure}[t]
 \centering
 \includegraphics[width=0.6\columnwidth,trim={0cm 8cm 1cm 8cm},clip]{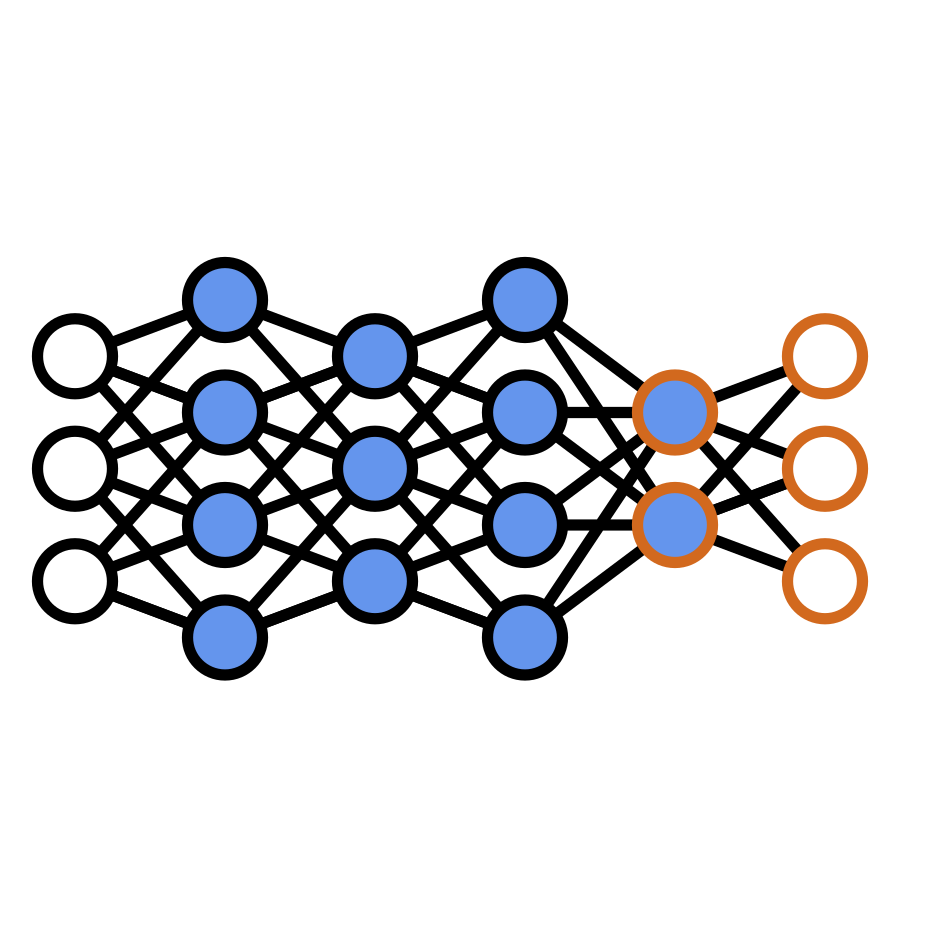}
 \caption{Schematic \emph{A2M2E} (inherently interpretable) model with bottleneck (depicted as 2 neurons). During training, bottleneck is (optionally) optimised to predict mid-level features, and based on this, output is optimised to predict emotions.}
 \label{fig:a2m2e}
\end{figure}

\textbf{A2M2E.} The second variant we look at was originally proposed as ``A2Mid2E-Joint''~\cite{Chowdhury2019MidLevel} and will be called ``A2M2E'' for brevity in what follows. This architecture is the same as for A2E except the last linear (classification) layer. Instead of this, a linear bottleneck layer is introduced, with the number of neurons corresponding to the number of mid-level features we would like to learn (here: seven). During training, this bottleneck layer is optimised to predict mid-level features (with an \gls*{mse} loss on the Mid-Level Features dataset), and its output is fed into a second linear layer, optimised to predict (eight) emotion annotations (with \gls*{mse} on the Soundtracks dataset). More precisely, to train the A2M2E model, both the predictions on mid-level features and on emotions are optimised simultaneously. This is done by first scaling the two losses (with a factor 0.5), and minimising their sum. In the following, the A2M2E model will be our inherently interpretable model, as it is designed to have a linear mapping from mid-level feature predictions --- which are assumed to be human-interpretable --- to final emotion predictions. A schematic view of this kind of network is shown in \figref{fig:a2m2e}.

\textbf{A2B2E.} Next to the A2M2E bottleneck model, we train a second model with the same (bottleneck) architecture, as depicted in \figref{fig:a2m2e}. For this second bottleneck model, we leave the bottleneck architecture as is, but remove the loss for optimising the mid-level features, and only learn to predict the emotion annotations. This model will be called \emph{A2B2E} in what follows, and is again a traditional, black-box model (as it directly predicts emotions based on input spectrograms) trained only on the Soundtracks dataset, yet with the same architecture as the interpretable A2M2E model. With this model, we will try to rule out any differences in robustness between our black-box and inherently interpretable model(s) solely being due to differences in their architectures (as A2E and A2M2E differ slightly). 

\subsection{Attacking the Emotion Recognition System}
\label{subsubsec:Emotion Classifer:Attack}
To test the robustness of the previously described models, we utilise adversarial attacks at test time. These allow us to compute marginal perturbations for input data, and determine how far they can change the output of a trained system. Specifically, we can assess the robustness or vulnerability of a model by looking at the performance of the model on the test inputs after completed training (i.e., performance \emph{before} an attack), computing perturbations for the test data, and comparing the performance before the attack to the performance on the modified test inputs (i.e., performance \emph{after} the attack).

In contrast to previous adversarial attacks in \gls*{mir}, where mostly classification systems were targeted (e.g., \cite{Kereliuk2015AudioAdversaries,Prinz2021Tismir}), the models at hand perform multiple regression tasks by predicting various emotion annotations. This requires us to rephrase previously introduced attacks. In particular, we formulate an untargeted attack, meaning that instead of changing the desired output of a model from one prediction to another (i.e., a target), we want to get as far away as possible from the available ground-truth emotion annotations. At the same time, the perturbed samples are restricted to lie within an $\epsilon-$ball around the original samples, so that we can ensure a certain degree of similarity. 

For the untargeted attack, we adapt \gls*{bim}~\cite{Kurakin2017BIM}. This method was introduced as an iterative variant of \gls*{fgsm}~\cite{Goodfellow2015FGSM}, and is also similar to a method that was introduced as \gls*{pgd} attack~\cite{Madry2018AdversarialTraining} (with one of the main differences being a random vs. zero initialisation for the perturbation). Intuitively, \gls*{bim} computes an adversarial perturbation by iteratively going into the gradient direction that maximises a model's loss w.r.t.~the ground-truth. More precisely, let $f$ be a model (e.g., our emotion recognition system), and $L$ the loss function it was trained with (e.g.,~\gls*{mse}). Then, let us denote an input by $\boldsymbol{x} \in \Reals^{S}$, the ground-truth (emotion) annotation by $\boldsymbol{y} \in \Reals^Y$, and the additive adversarial perturbation by $\boldsymbol{\delta} \in \Reals^{S}$ (i.e.,~making an adversarial example $\hat{\boldsymbol{x}} = \boldsymbol{x} + \boldsymbol{\delta}$). We compute $\boldsymbol{\delta}_{ep+1}$, i.e., the perturbation of the next iteration, as
\begin{equation}
    \boldsymbol{\delta}_{ep+1} = \clip_\epsilon (\boldsymbol{\delta}_{ep} + \eta * \sign(\nabla_{\boldsymbol{\delta}_{ep}} \ L(f(\boldsymbol{x} + \boldsymbol{\delta}_{ep}), \boldsymbol{y}))).
\end{equation}

\noindent
Here, $\nabla_{\boldsymbol{\delta}}$ is the gradient w.r.t.~$\boldsymbol{\delta}$, and updates are scaled with multiplicative factor $\eta$ and performed based on the $\sign$ of the gradient. Furthermore, we $\clip$ each perturbation to stay in the range of $[-\epsilon, \epsilon]$. We initialise $\boldsymbol{\delta}_{0} = \boldsymbol{0}$. 

This adapted regression attack is iterative, i.e., we perform updates to $\boldsymbol{\delta}$ multiple times. We do so, until 1)~a number of maximum iterations is carried out, or 2)~we reach a certain threshold either on the current \gls*{mse} --- which is also the loss function used during training of the model --- or on the average correlation between prediction and target emotion. More specifically, we perform experiments with both the \gls*{mse} as well as the correlation threshold, and stop updating $\boldsymbol{\delta}$ ahead of the maximum number of iterations if a pre-defined upper / lower limit is reached.

\subsection{Adversarial Training to Improve Robustness}
\label{subsec:Adversarial Training}
Adversarial training is one of the most prominent methods to increase a deep learning model's robustness~\cite{Madry2018AdversarialTraining,Zhang2019YOPO,Shafahi2019FreeAdversarialTraining}. The main idea of adversarial training is to include adversarial examples in the training data, to decrease a model's sensitivity against small alterations of clean input data (i.e., adversarial perturbations). More precisely, for a network $f$ defined via the parameters $\theta$ and trained with loss $L$, adversarial training aims at solving the saddle point problem~\cite{Madry2018AdversarialTraining}
\begin{equation}
    \min_\theta \ \Expect_{(\boldsymbol{x}, \boldsymbol{y}) \sim \cD} \left[ \max_{\boldsymbol{\delta} \in B(\boldsymbol{x}, \epsilon)} \ L(f(\boldsymbol{x} + \boldsymbol{\delta}), \boldsymbol{y}) \right]. 
\end{equation}

Here, $(\boldsymbol{x}, \boldsymbol{y}) \sim \cD$ denotes the data that a system $f$ is trained on, and $B(\boldsymbol{x}, \epsilon)$ is the set of allowed adversarial perturbations $\boldsymbol{\delta}$ for a data point $\boldsymbol{x}$, i.e., $B(\boldsymbol{x}, \epsilon) = \{\boldsymbol{\delta} \in \Reals^S \mid \norm{\boldsymbol{\delta}}_p \leq \epsilon\}$. Intuitively, this means we train a model to have a low loss on the training data, which can, however, be perturbed by an adversarial attack before it is fed into the system.

In what follows, we obtain our robust models by optimising on a clean batch first, perturbing said batch by means of an adversarial attack, and then optimising on the according adversarial batch. We do so every $n^{th}$ epoch of the training process. In our experiments, the perturbations for adversarial training are computed with the adapted \gls*{bim} method. We use ``aA2E'' and ``aA2B2E'' to denote the robust version of models A2E and A2B2E, respectively. 

\section{Experiments}
\label{sec:Experiments}
In this section, we first describe our overall experimental setup and the performance of previously discussed emotion recognition systems. Thereafter, we present and discuss our main findings. To support the reproduction of the following experiments, the code is available via Github.\footnote{\url{https://github.com/CPJKU/robustness_mer_bottleneck}}

\subsection{Experimental Setup}
\label{subsubsec:Emotion Recognition:Experimental Setup}
In the subsequent experiments, we follow Chowdhury et al.~when training the emotion recognition systems~\cite{Chowdhury2019MidLevel}, with minor adaptions. We use a training, validation, and a test-set with 80\%-10\%-10\% of the data (in contrast to Chowdhury et al., who only use a training and validation-set~\cite{Chowdhury2019MidLevel}). We do this to have a better approximation of the final model performance~\cite{Hastie2009StatisticalLearning}. As we adopt the hyperparameters used in the original work~\cite{Chowdhury2019MidLevel}, we refrain from using e.g., cross-validation to estimate the prediction error for hyperparameter tuning, and instead use a fixed training and validation-set. We repeat the training of the models 10 times for different random initialisations of the networks. Furthermore, we take Adam as an optimiser to minimise the loss (or combined loss, in case of A2M2E) for training, and a learning rate of 0.0005~\cite{Chowdhury2019MidLevel}. The batch size for training is 8, and we train for at most 200 epochs, with early stopping being used against overfitting (patience of 50).

To find suitable hyperparameters for the adversarial attack on the system, we perform a grid-search on the validation data. We choose a hyperparameter setting that trades off a high \gls*{mae} and low correlation between annotations and labels after an attack. This leads to a number of maximal iterations of $1,000$, $\epsilon = 0.001$, $\eta = 0.002$, and a lower correlation threshold of -1. 
A repetition of the attack on one particular model (for multiple random seeds) is not necessary; due to the lack of a source of randomness for the adversarial attack (e.g., randomly sampled targets), the attack is deterministic. 

In the case of adversarial training, we opt to train on adversarial batches every five epochs, which in preliminary experiments led to a suitable trade-off between performance on clean data, and adversarial robustness. We use the same attack parameters as for the attack on the systems, yet with a maximal number of 50 iterations. 

Regarding the evaluation of model performance, we look at two measures, computed on the test-set. First, we state the correlation between emotion annotations and model predictions, as was done by Chowdhury et al.~\cite{Chowdhury2019MidLevel}, to determine the performance of the emotion recognition models. Additionally, we have a look at the \gls*{mae}, as we cannot guarantee a linear relation between emotion annotations and predictions (particularly on adversarial samples; see also Section~\ref{subsec:Looking At Predictions}), making \gls*{mae} a better suited metric to capture the performance (cf.~\cite{Flexer2021InterIntraAgreement}).

\subsection{Emotion Recognition Results}
\label{subsubsec:Reproduction:Emotion Recognition}
We now look at the performance of our reproduction on the task of emotion prediction on the Soundtracks dataset. In Table~\ref{tab:emotion_reproduction}, we summarise the performance with Pearson’s correlation coefficient, as done in the original work~\cite{Chowdhury2019MidLevel}. Specifically, we list the overall average (i.e., over 8 targets and 10 runs) correlation between emotion annotations and predictions. The first two entries of the table show the performance reported by Chowdhury et al.~\cite{Chowdhury2019MidLevel}, as the average over ten random runs (i.e., running on 10 different training and validation splits). Entries 3-5 in Table~\ref{tab:emotion_reproduction} show our reproduction results as the average and standard deviation over ten runs, with different random initialisations of the networks.
We also add the performance of adversarially trained variants of the previous networks in the final two entries of the table. Our results are reported on the test-set.

\begin{table}
 \begin{center}
 \begin{tabular}{l c c}
  \toprule
  System & Avg Corr. & MAE \\
  \midrule
  A2E (orig.) & 0.76* & - \\
  A2M2E (orig.) & 0.75* & - \\
  \midrule
  A2E & 0.73 $\pm$ 0.02 & 0.10 $\pm$ 0.01 \\ 
  A2B2E & 0.70 $\pm$ 0.02 & 0.11 $\pm$ 0.01 \\ 
  A2M2E & 0.67 $\pm$ 0.02 & 0.11 $\pm$ 0.01 \\ 
  \midrule
  aA2E & 0.67 $\pm$ 0.03 & 0.10 $\pm$ 0.00 \\ 
  aA2B2E & 0.66 $\pm$ 0.03 & 0.10 $\pm$ 0.00 \\ 
  \bottomrule
 \end{tabular}
\end{center}
 \caption[Performance of emotion recognition models.]{Performance of emotion recognition models reported by \cite{Chowdhury2019MidLevel} (orig.) and our reproduction, as well as two adversarially trained models. Our results given as mean $\pm$ standard deviation over 10 runs on test data. Note that original results (*) are computed on a validation-set, while ours are reported on a test-set.}
 \label{tab:emotion_reproduction}
\end{table}

We achieve almost on-par performance of the A2E model compared to the original work~\cite{Chowdhury2019MidLevel} for the average label correlation (difference of 0.03). However, for the interpretable A2M2E variant, this difference is larger (0.08), which can likely be attributed to our performance estimate being on a hold-out test-set, as opposed to a (more biased) cross-validation estimate (i.e., on validation-sets).\footnote{Note that when recreating the data-splits in \cite{Chowdhury2019MidLevel}, we achieve higher performing models, and a notably larger standard deviation between different runs (e.g., A2E: $0.88 \pm 0.07$, A2M2E: $0.80 \pm 0.06$), supporting this assumption.} 

In addition to the label correlation, we also look at the \gls*{mae} as a performance measure. The \gls*{mae} is, across all our standard training settings, between 0.10 and 0.11, with low standard deviation for different runs. This performance will be used as a point of reference to measure the robustness of the models in subsequent experiments. 

In the last two entries of Table~\ref{tab:emotion_reproduction} we show the results of adversarially trained versions of the black-box models A2E and A2B2E. Similar to what was shown previously~\cite{Ilyas2019AdvBugs}, we observe a slightly reduced performance of the more robust models on the clean test-set, in terms of average correlation, compared to the black-box variants. For the \gls*{mae}, the performance remains at 0.10, and a low (here: 0.00) standard deviation across the ten random runs. 

\subsection{Robustness of Black-Box vs.~Interpretable Models}
\label{subsec:Emotion Recognition:Results}
After looking at the performance of the different types of models (i.e., black-box, adversarially trained, and inherently interpretable) in the previous section, we now look at their robustness against adversarial attacks, by computing the performance after such an attack (i.e., on the modified test data). In our experiments, we look at the \gls*{mae} of the models. We expect robust models to remain at a lower \gls*{mae}, and less robust models to exhibit higher \glspl*{mae} after an attack than before (i.e., on the clean test data).

\begin{figure}[t]
    \centering
    \includegraphics[width=1\columnwidth,trim={2.5cm 0cm 3.5cm 0cm},clip]{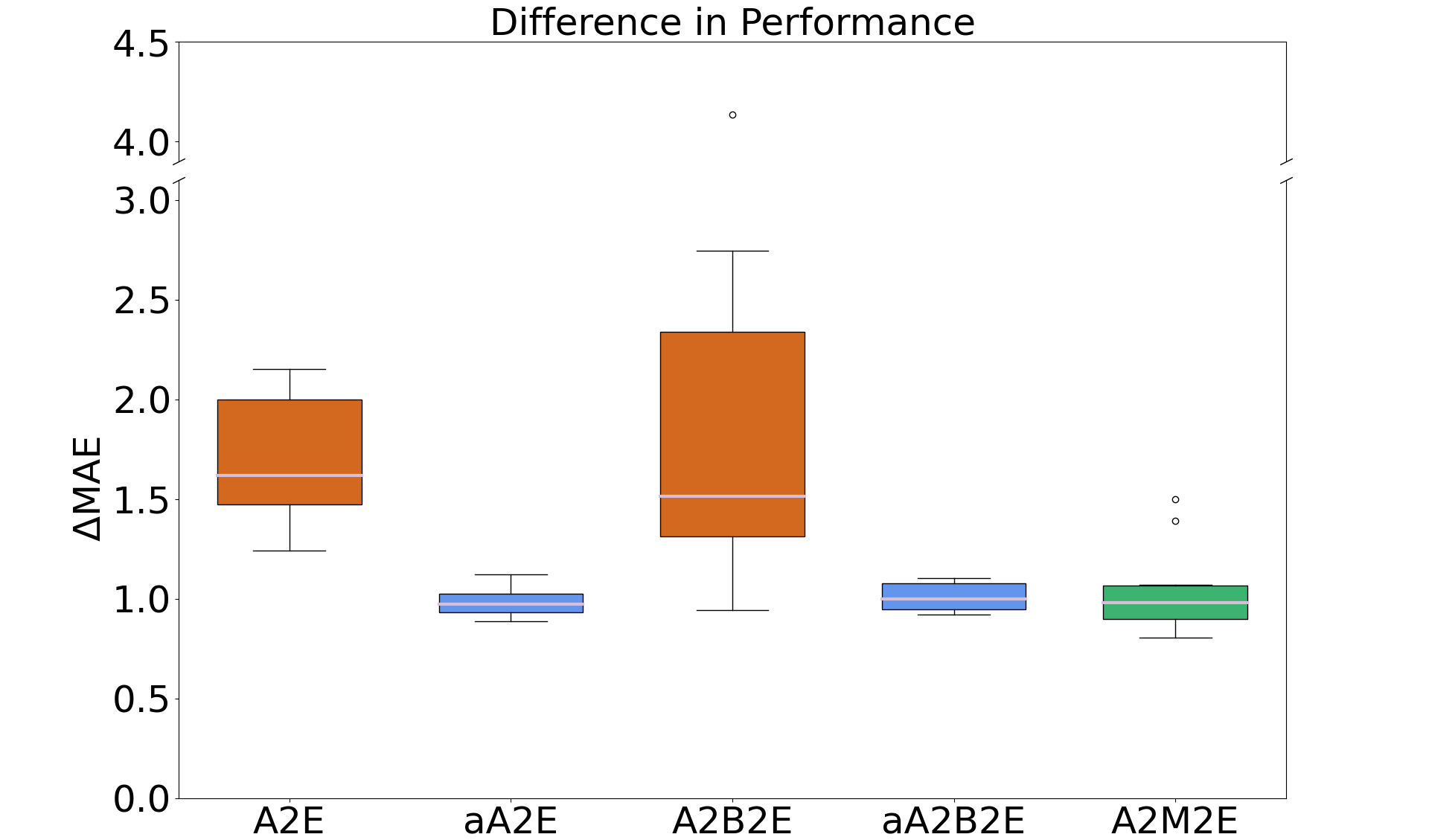}
    \caption{Model robustness, illustrated by impact of adversarial attack. Depicted is difference in performance after attack vs.~before attack ($\Delta$MAE), for different models. 
    Lower $\Delta$MAE indicates higher robustness. Orange box-plots (A2E and A2B2E) are black-box models, blue box-plots (aA2E, aA2B2E) are adversarially trained models; green (A2M2E) is the inherently interpretable model (without any adversarial training). The plots are based on ten random runs.}
    \label{fig:performance_difference}
\end{figure}

In Figure~\ref{fig:performance_difference}, we show the difference in performance of our models before and after an attack, subsequently denoted by $\Delta$MAE. Note here that the performance before an attack is rather similar for all models (Table~\ref{tab:emotion_reproduction}). The $\Delta$\gls*{mae} is computed across the ten random runs again. The first and third box-plot represent the losses in performance for the black-box models; the second and fourth box-plots for their respective adversarially trained versions. Finally, the last box-plot depicts the losses in performance for the inherently interpretable model across ten runs.

Figure~\ref{fig:performance_difference} shows, first of all, that the two black-box models A2E and A2B2E have the highest (median) loss in performance. They also exhibit the largest variations between different training runs, yet with a clear tendency to be more susceptible to input perturbations. The figure also shows the benefit of including adversarial examples during training, as the two adversarially trained models aA2E and aA2B2E appear distinctly more robust than their black-box counterparts. Also the variations across different runs are lower. Remarkably enough, the performance (loss) of the inherently interpretable model, shown in the last box-plot of Figure~\ref{fig:performance_difference}, is much more similar to the adversarially trained networks. It tends to have much lower \gls*{mae} than the black-box models after an attack, without the need for a supplementary addition of adversarially perturbed data during training. Note that preliminary experiments show, that the robustness of the concept bottleneck model could be further improved via adversarial training, yet with an even greater loss in performance on clean data.

We also run a two-sided t-test with a Bonferroni correction to validate the experiments, using a significance threshold of $\alpha= 0.05/2$ (correcting for two hypothesis tests). This test shows that the difference in robustness between both black-box models and the interpretable model is significant (A2E vs.\ A2M2E: t(359)=13.04, p=0.00; A2B2E vs.\ A2M2E: t(359)=12.10, p=0.00).

Note that another aspect of adversarial perturbations, namely their perceptibility, is largely neglected here, except for the restrictions of staying within an $\epsilon$-ball around the original samples. This is done as our main focus is to determine the overall susceptibility of the models to small input perturbations. However, looking at the approximate signal-to-noise ratio over all perturbed samples, we find comparable levels of perceptibility across the different models ($\approx32-37$dB, cf.~\cite{Prinz2021Tismir}, where samples with around 40dB were moderately perceptible). 

\subsection{A Closer Look at the Predictions}
\label{subsec:Looking At Predictions}

\begin{figure*}[t]
    \centering
    \includegraphics[width=0.85\linewidth]{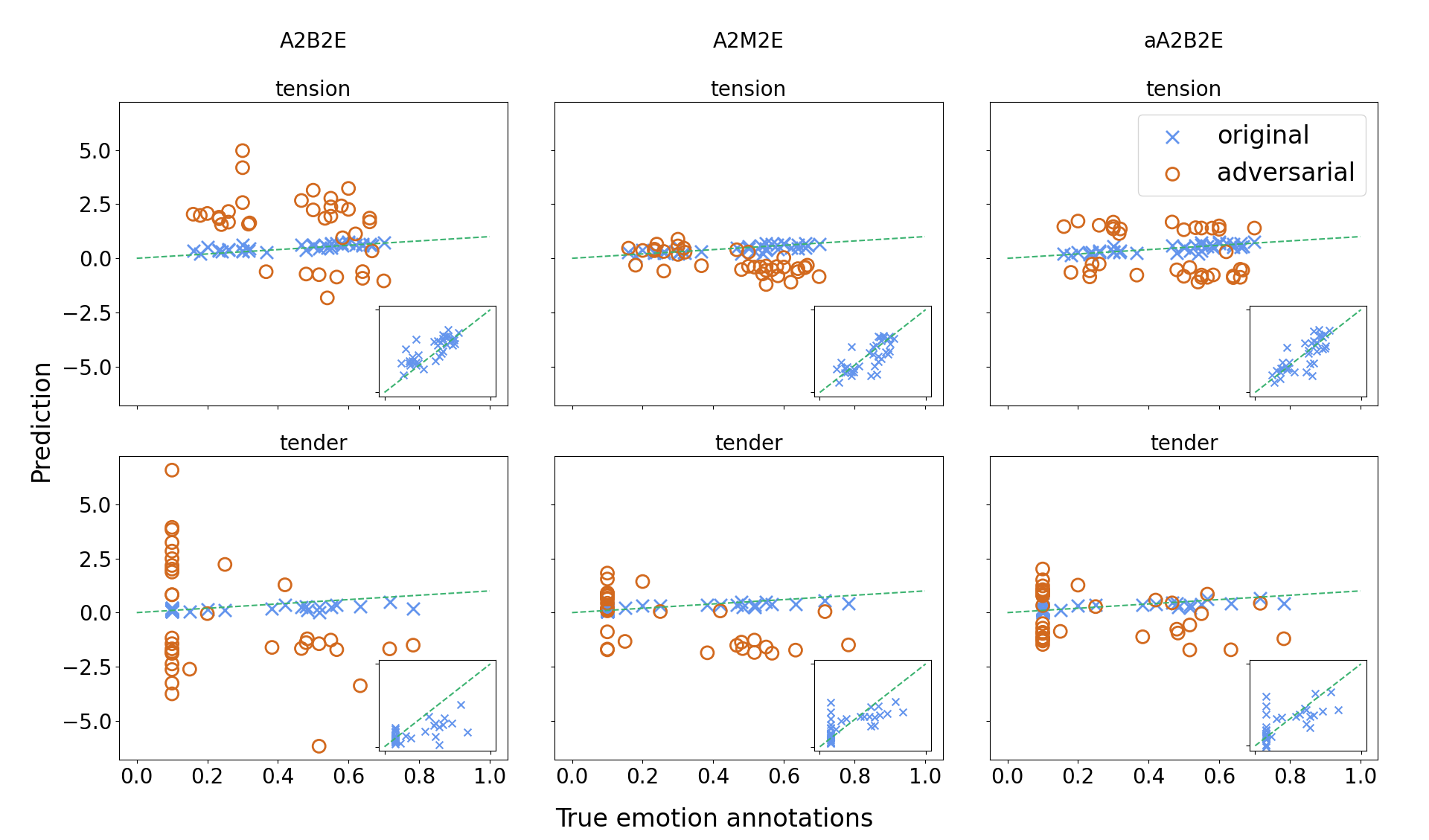}
    \caption{True emotion annotation (x-axis) vs. predictions (y-axis) for two different emotions (tension and tender), and three different models. Leftmost plot (A2B2E) presents predictions of black-box model, middle column (A2M2E) inherently interpretable model, and rightmost plot (aA2B2E) shows predictions of adversarially trained model. Depicted are predictions on original (clean) data (``x''), and adversarially perturbed data (``o'').}
    \label{fig:predictions_tension_tender}
\end{figure*}

In this section, we look more closely at some individual predictions made by our models. Figure~\ref{fig:predictions_tension_tender} contrasts true emotion annotations (x-axis) and predictions (y-axis) of different models, for two of the eight predicted emotions (``tension'' and ``tender''). Depicted are predictions both on original, clean test data (``x''), and on adversarially perturbed data (``o''). Note that the ground truth emotion annotations (horizontal axes) are strictly confined to the range $[0,1]$; the emotion recognition model that we are using~\cite{Chowdhury2019MidLevel}, however, as a regression model, does not restrict the predicted numeric output to any particular range.  
Zooming in on the $[0,1]$ range on the y-axis (see inserts in the individual plots), we see that the models' predictions for the original data stay well within the desired range, while the perturbed samples lead the models to output wildly different numbers. 

Moreover, while the original predictions appear to have mostly linear relationships with the ground-truth annotations, the adversarial predictions exhibit less thereof, justifying our choice of using the \gls*{mae} instead of the correlation when comparing the robustness of different models.

From Figure~\ref{fig:predictions_tension_tender}, we also see that for the emotions of tension and tender, for the black-box model A2B2E (left), the predictions on adversarially perturbed data appear a lot further from the original predictions than for both the interpretable model (A2M2E, middle) and the adversarially trained version (aA2B2E, right), which again illustrates the latter models' higher robustness.
Interestingly, for emotions with a lot of very low ground-truth annotations (e.g., tender), the adversarial perturbations tend to lead to widely spread out predictions.
We also note that for other emotions (e.g., anger, energy) predictions on adversarial data look more similar across different models, suggesting that different emotions are easier / harder to perturb. 

\section{Conclusion and Discussion}
\label{sec:Conclusion and Discussion}
In this paper, we investigated whether an inherently more interpretable system could provide a more robust solution to \gls*{mir} tasks. We looked at an emotion recognition system designed to encode its predictions in terms of human-understandable concepts, and compared its robustness against adversarial input perturbations to the robustness of a traditional, black-box model. We found that this kind of concept bottleneck model can present a more robust alternative to a black-box model. In fact, it can even exhibit robustness similar to adversarially trained models --- which are specifically trained to be more robust against adversarial input alterations --- yet without the cost of needing to compute adversarial examples during training. 

An obvious limitation of this set of experiments is that it considered only one single model and recognition task so far. This calls for broader and more systematic analyses to further support the proposed hypothesis, and provide more insights into the complex connection between interpretability and robustness. As there has only been a limited amount of research in MIR on interpretable models (e.g., \cite{Chowdhury2019MidLevel,Zinemanas2021APNet,Loiseau2022PlayablePrototypes}), we additionally hope to see more work in this direction.

\begin{acknowledgments}
This research was funded in part by the Austrian Science Fund (FWF) [10.55776/AR821 and 10.55776/P36653]. GW's work is supported by the European Research Council (ERC) under the EU's Horizon 2020 research and innovation programme, grant agreement No 101019375 (\textit{Whither Music?}).
\end{acknowledgments} 
	
\bibliography{smc2025}
	
\end{document}